\documentclass[11pt]{article}

\usepackage{times}

\usepackage{graphicx}

\usepackage{amsmath}

\newenvironment{sciabstract}{%
\begin{quote} \bf}
{\end{quote}}

\title{Melting a skyrmion lattice topologically: through the hexatic phase to a skyrmion liquid }

\author
{Ping Huang,$^{1,2\ast}$ Marco Cantoni,$^{3}$ Arnaud Magrez,$^{4}$ Fabrizio Carbone,$^{2}$ Henrik M. R{\o}nnow $^{1\ast}$ \\
\\
\normalsize{$^{1}$Laboratory for Quantum Magnetism (LQM), Institute of Physics,}\\
\normalsize{\'{E}cole Polytechnique F\'{e}d\'{e}rale de Lausanne (EPFL), CH-1015 Lausanne, Switzerland}\\
\normalsize{$^{2}$Laboratory for Ultrafast Microscopy and Electron Scattering (LUMES), Institute of Physics,}\\
\normalsize{\'{E}cole Polytechnique F\'{e}d\'{e}rale de Lausanne (EPFL), CH-1015 Lausanne, Switzerland}\\
\normalsize{$^{3}$Centre Interdisciplinaire de Microscopie \'{E}lectronique (CIME),}\\
\normalsize{\'{E}cole Polytechnique F\'{e}d\'{e}rale de Lausanne (EPFL), CH-1015 Lausanne, Switzerland}\\
\normalsize{$^{4}$Crystal Growth Facility, Institute of Physics,}\\
\normalsize{\'{E}cole Polytechnique F\'{e}d\'{e}rale de Lausanne (EPFL), CH-1015 Lausanne, Switzerland}\\
\normalsize{$^\ast$To whom correspondence should be addressed;} \\
\normalsize{E-mail: ping.huang@epfl.ch; henrik.ronnow@epfl.ch}
}

\date{}

\begin{document}

\baselineskip18pt

\maketitle

\begin{sciabstract}

Skyrmions are twirling magnetic textures whose non-trivial topology leads to particle-like properties promising for information technology applications. Perhaps the most important aspect of interacting particles is their ability to form thermodynamically distinct phases from gases and liquids to crystalline solids. Dilute gases of skyrmions have been realized in artificial multilayers, and solid crystalline skyrmion lattices have been observed in bulk skyrmion hosting materials. Yet, to date melting of the skyrmion lattice into a skyrmion liquid has not been reported experimentally. Through direct imaging with cryo-Lorentz transmission electron microscopy, we demonstrate that the skyrmion lattice in the material Cu$_\mathbf{2}$OSeO$_\mathbf{3}$ can be dynamically melted. Remarkably, we discover this melting process to be a topological defects mediated two-step transition via a theoretically hypothesized hexatic phase to the liquid phase. The existence of hexatic and liquid phases instead of a simple fading of the local magnetic moments upon thermal excitations implies that even in bulk materials skyrmions possess considerable particle nature, which is a pre-requisite for application schemes.

\end{sciabstract}

Magnetic skyrmions\cite{bogdanov_thermodynamically_1989} have been attracting increasing attention due to their promising potential to be the building block for the next generation spintronics\cite{fert_skyrmions_2013} originating from their novel properties\cite{nagaosa_topological_2013} and especially their unique topology\cite{braun_topological_2012}. Among various skyrmion hosting materials, bulk crystalline compounds such as MnSi and Cu$_2$OSeO$_3$ host a skyrmion lattice (SkL) phase in which skyrmions form a two-dimensional (2D) crystal. One of the most important issues that should be addressed is whether and how a SkL ``melts''. There have long been two pictures in describing a SkL: assemblies of quasi-particles\cite{karube_robust_2016} or superposition of coherent spin helices\cite{muhlbauer_skyrmion_2009}. In the former scenario, individual skyrmions should persist throughout the whole melting process until an atomic level magnetic phase transition happens (\emph{e.g.} transforming to the paramagnetic phase or the spin polarized state), with the long range orders evolving accordingly\cite{timm_skyrmion_1998}. While in the latter description, the phase coherence among the spin helices will be destroyed by thermal fluctuations, leading to the vanishing of skyrmions, without an intermediate processes of lattice symmetry lifting. Thus in the latter case a SkL ``fades'' rather than melts\cite{ambrose_melting_2013}.

If a SkL does melt, its 2D nature may give rise to non-trivial phase behaviors. Generally, unlike in three dimension, there is no true long range positional order in 2D solids due to strong fluctuations\cite{mermin_crystalline_1968}, but long range orientational order can still exist. According to the Kosterlitz, Thouless, Halperin, Nelson, and Young (KTHNY) melting theory\cite{kosterlitz_ordering_1973,nelson_dislocation-mediated_1979,young_melting_1979}, these two types of orders can evolve separately due to the disassociation of topological defects. This leads to a phase evolution from the solid phase, through a unique intermediate phase, the \emph{hexatic} phase, finally to the liquid phase. The hexatic phase possesses short range positional and quasi-long range orientational orders while the latter will further be broken in the liquid phase. Although previous theoretical analysis predicted such a two-step KTHNY melting process in SkLs upon increasing temperature\cite{timm_skyrmion_1998}, recent Monte Carlo simulation studies did not find any sign of the hexatic phase\cite{ambrose_melting_2013,nishikawa_solid--liquid_2017}. To date, direct experimental studies have been lacking.

Here, we use real space cryo-Lorentz Transmission Electron Microscopy (LTEM) to quantify the melting of the SkL in a 150 nm thin single crystalline slab of insulating skyrmion hosting compound Cu$_2$OSeO$_3$\cite{scnmm_2018}. We discover that the spatial and temporal evolutions of the SkL correlations upon the increase of the magnetic field can be described by a two-step melting phase transition through an intermediate hexatic phase to a liquid phase, as summarized in the schematic phase diagram shown in Fig. 1. In contrast, only the skyrmion lattice phase has been addressed in literature\cite{seki_formation_2012}.

\begin{figure*}
  \centering
  \includegraphics{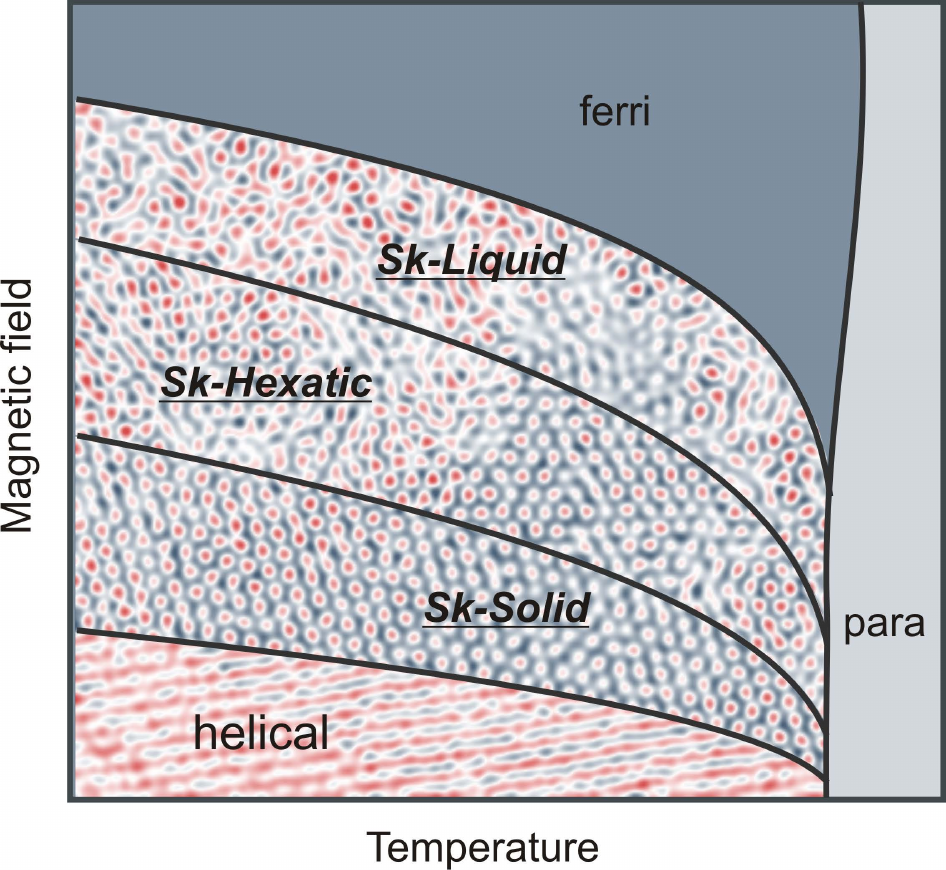}
  \caption{\textsf{\textbf{Schematic phase diagram of nano-slab Cu$_2$OSeO$_3$.} Compared to the previously reported results, where rough phase classification included only the helical, the SkL and the ferri-phases, our real space LTEM investigations reveal both a hexatic and a liquid phase.}}
\end{figure*}

Real space LTEM images at selected magnetic fields and at a constant temperature $T=28$ K are shown in Fig. 2, A to C. Their corresponding Fourier transforms (FT) are shown in inserts, exhibiting three typical types of arrangements of the skyrmions. At low magnetic field, $H=444$ Oe (Fig. 2A), the conventional triangular SkL commonly reported in literature\cite{seki_observation_2012} is observed. Its FT are six sharp Bragg peaks forming a regular hexagon, similar to the small angle neutron scattering (SANS) patterns reported previously\cite{muhlbauer_skyrmion_2009,adams_long-wavelength_2012,seki_formation_2012,white_electric-field-induced_2014}. Those sharp Bragg peaks evolve into arcs at higher magnetic field, as shown in Fig. 2B for $H=887$ Oe. Eventually, at a high enough magnetic field, $H=1124$ Oe, the skyrmions are in an isotropic state, as indicated by the circular FT pattern shown in Fig. 2C.

\begin{figure*}
  \centering
  \includegraphics{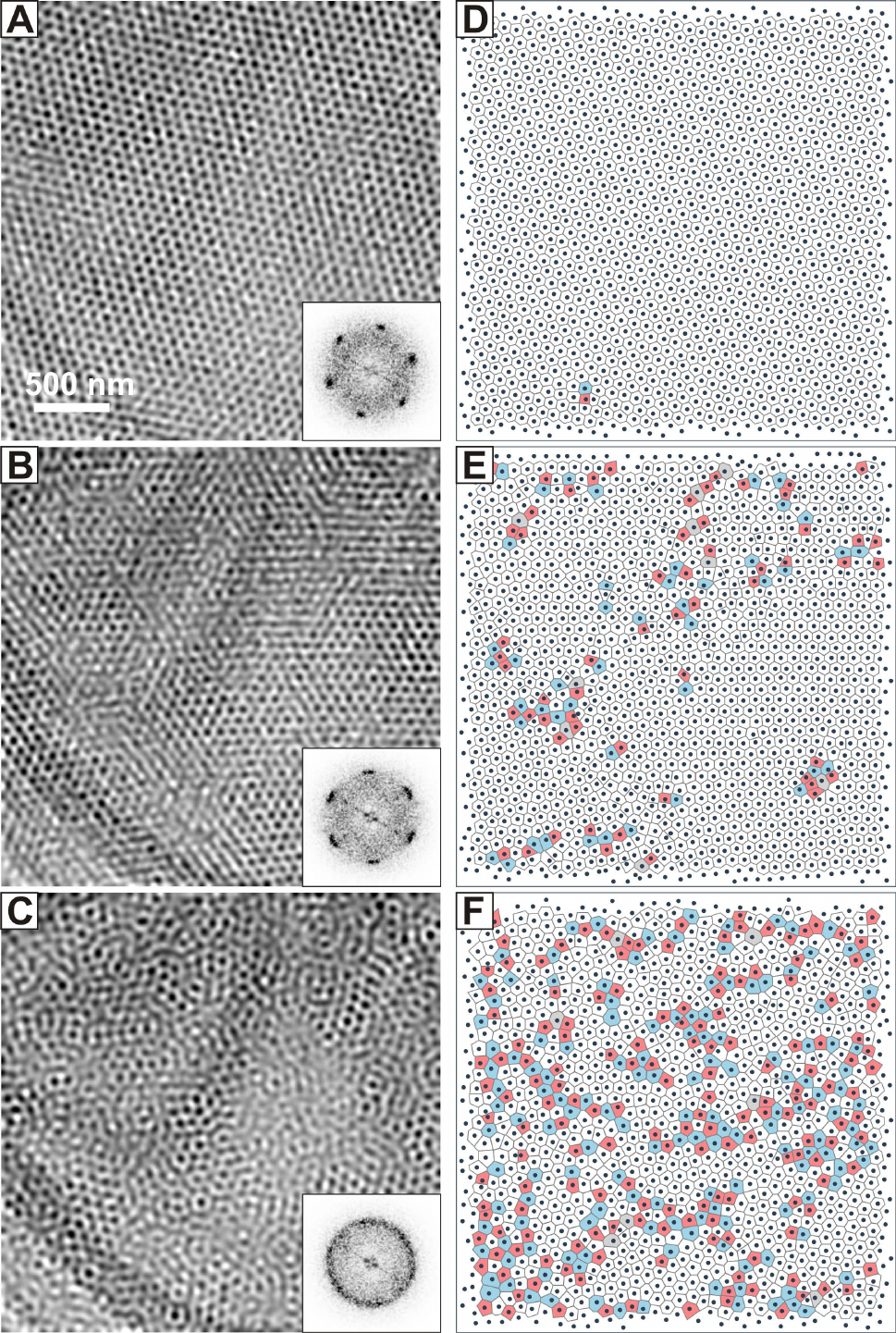}
  \caption{\textsf{\textbf{Real space results and analysis of the SkL melting process.} Real space LTEM images of the SkL at the magnetic field of (\textbf{A}) 444 Oe, (\textbf{B}) 887 Oe and (\textbf{C}) 1124 Oe. Inserts are the corresponding FT. (\textbf{D} to \textbf{F}) The solid dots denote the positions of the skyrmions in (\textbf{A}) to (\textbf{C}) respectively. The Voronoi tessellation is carried out for each assembly of skyrmions. Red, blue and gray polygons indicate sites with 5, 7 and more than 7 neighbors respectively, and hexagons are left blank.}}
\end{figure*}

Further understanding of the melting process of the SkL in real space can be achieved by topology analysis. An algorithm was used to identify all skyrmions in the image data\cite{scnmm_2018}, as shown in Fig. 2, D to F, corresponding to Fig. 2, A to C respectively, where solid dots represent the skyrmion positions. Voronoi tessellation based on the skyrmion positions is then performed and shown as edge-sharing polygons in Fig. 2, D to F. In a 2D triangular lattice, every site topologically has 6 neighbors forming a regular hexagon, and non-6 sites can form topological defects, \emph{i.e.} dislocations and disclinations. A dislocation is a pair of 5- and 7- sites, and a disclination is a single 5- or 7- site. It can be clearly seen that as the magnetic field increases, the density of the topological defects increases correspondingly, as shown quantitatively in Fig. 4B, implying that the melting process is mediated by the generation of topological defects upon the increase of the magnetic field.

Calculating spatial and temporal correlation functions from the skyrmion positions provides quantitative insights. The pair correlation function \cite{lu_applications_1997} characterizes the possibility of finding a particle at a given distance from a reference point, thus reflecting the positional correlation of an assembly of particles:

\begin{equation}
  G_r(r) = \frac{1}{\pi N r \rho_0} \sum_{j=1}^{N} \sum_{i > j}^{N} \delta (r-r_{ij})
  \label{eq:eq1}
\end{equation}
where $N$ is the total number of the skyrmions and $\rho_0 = N/V$ is the average skyrmion density. Fig. 3A shows the pair correlation function $G_r(r)$ of different phases. In the solid phase, one can see sharp peaks at well defined position corresponding to a triangular lattice. These peaks can be seen up to the maximum distance determined by the field of view. However, in the hexatic phase, although peaks show up at similar positions, closely packed peaks merge into broad bumps. When the system enters the liquid phase, no correlations extend beyond 500 nm. This evolution of $G_r(r)$ with the increasing magnetic field reflects the gradual loss of positional order.

\begin{figure*}
  \centering
  \includegraphics{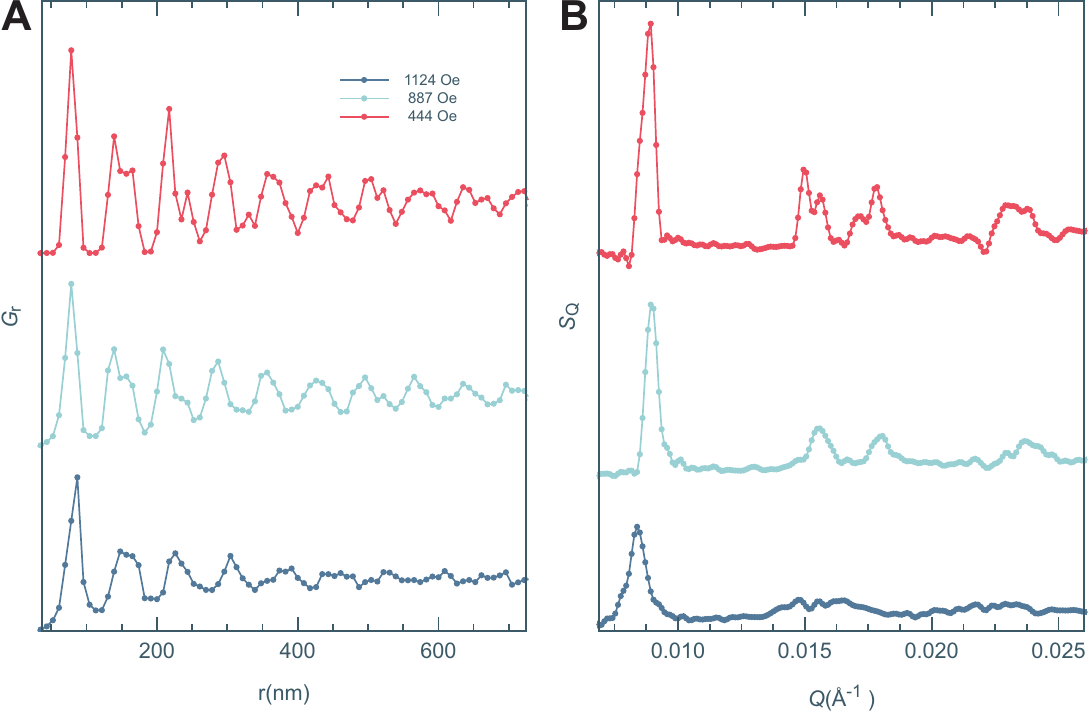}
  \caption{\textsf{\textbf{Positional order of the SkL at different magnetic fields.} (\textbf{A}) The pair correlation functions and (\textbf{B}) the structure factors of the SkL at different magnetic fields. Curves are shifted vertically for clarity.}}
\end{figure*}

From the pair correlation function $G_r(r)$, we can calculate the structure factor $S(Q)$\cite{lin_time-resolved_2006,scnmm_2018}, as shown in Fig 2B. In the solid phase, peaks are sharp and well resolved, reflecting a well organised periodic structure. In the hexatic phase, a significant reduction in peak amplitude can be observed. Similar to $G_r(r)$, closely spaced sharp peaks merge into broad ones. These peaks merge further into wide bumps or even vanish in the liquid phase. Note that an increase of the skyrmion spacing can be clearly evidenced by the positions of the primary peak which shifts from 0.00893 $\rm \AA^{-1}$ in the solid phase to 0.00839 $\rm \AA^{-1}$ in the liquid phase, indicating the skyrmion spacing to be 61.0 nm and 64.9 nm respectively for the solid and the liquid phase, consistent with theoretical analysis\cite{han_skyrmion_2010}.

For a 2D triangular lattice, the orientational order parameter can be defined through the orientations of the bonds between nearest neighbor sites\cite{nelson_dislocation-mediated_1979}:
\begin{equation}
  \begin{aligned}
    \Psi_6 (\mathbf{r}_i) &= \frac{1}{N_{nn}} \sum_{j=1}^{N_{nn}}  e^{i6\theta_{ij}}
  \end{aligned}
  \label{eq:eq4}
\end{equation}
where $N_{nn}$ represents the number of the nearest neighbouring particles around the reference particle located at position $\mathbf{r}_i$, which can be determined by Delaunay triangulation. $\theta_{ij}$ is the angle between the $i-j$ bond and an arbitrary but fixed axis. The orientational correlation function $G_6(r)$ is then:
\begin{eqnarray}
  \begin{aligned}
    G_6 (r) = \frac{1}{N_r} \sum_{\left< i,j \right>}^{N_r} \Psi_6 (\mathbf{r}_i) \Psi_6^* (\mathbf{r}_j)
  \end{aligned}
  \label{eq:eq3}
\end{eqnarray}
where $N_r$ is the number of particles at distance $r$ away from each other.

\begin{figure*}
  \centering
  \includegraphics{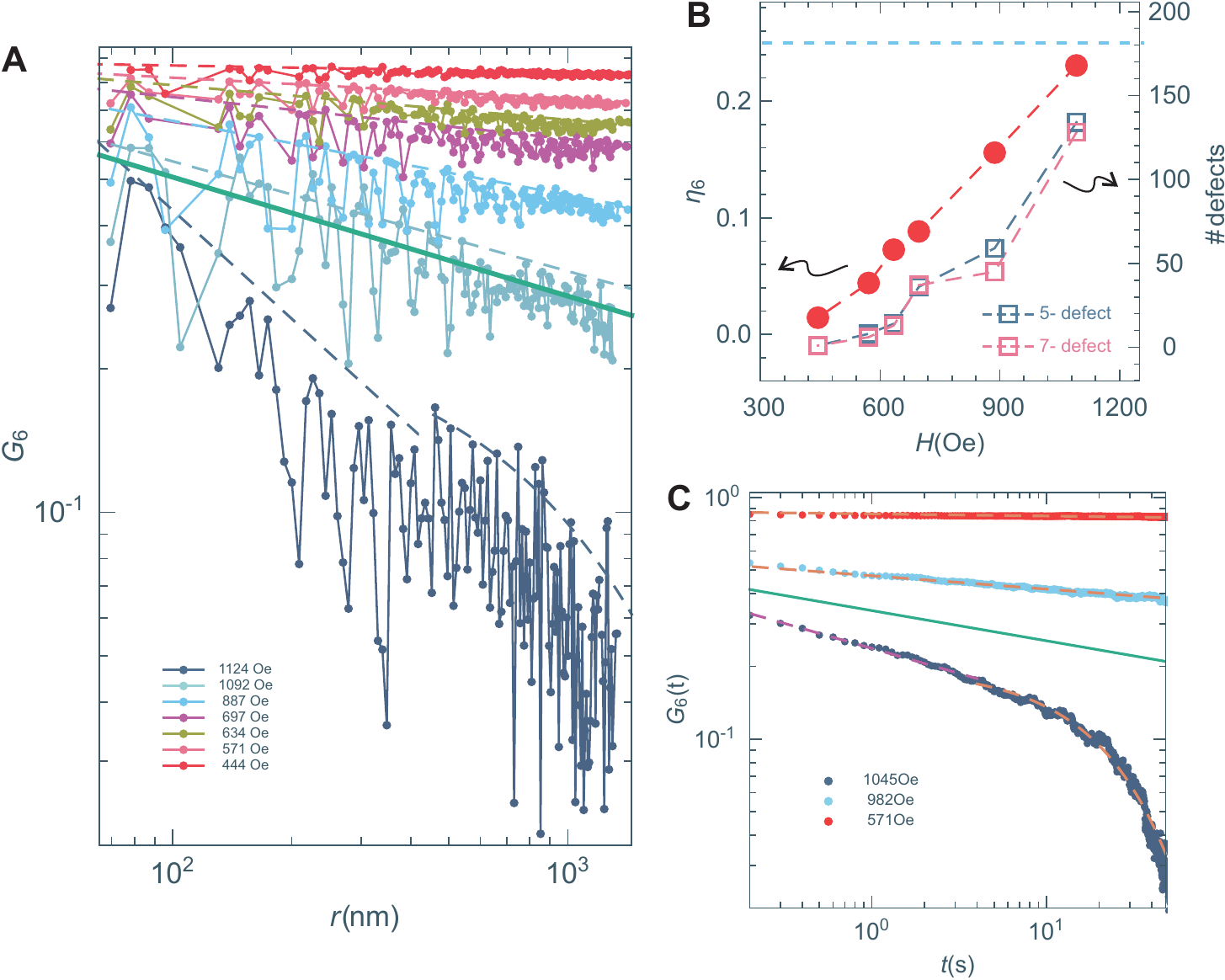}
  \caption{\textsf{\textbf{Orientational order of the SkL at different magnetic fields.} (\textbf{A}) The spatial orientational correlation functions $G_6(r)$ at different magnetic fields. The dashed lines are fits to the upper envelopes of the data by power-law or exponential decays respectively. Note that for $\mathsf{H=1124}$ Oe, exponential decay is used for fitting at large distances and power-law for short distances. The critical power law decay exponent 0.25 predicted by the KTHNY theory is illustrated by the solid green line. (\textbf{B}) The magnetic field dependence of the derived algebraic decay indexes $\mathsf{\eta_6}$. The dashed horizontal line denotes the critical value of $\mathsf{\eta_6 = 0.25}$ for the hexatic to the liquid phase transition predicted by the KTHNY theory. The numbers of \textit{5-} and \textit{7-}defects at each field are shown in empty squares respectively. (\textbf{C}) Temporal correlation functions of the orientational order parameter $G_6(t)$ culcuated from three LTEM image series acquired at $H=571$ Oe, $H=982$ Oe and $H=1045$ Oe, corresponding to the solid, the hexatic and the liquid phases respectively. The derived $G_6(t)$ are shown in solid dots in the $\log$-$\log$ coordinate system. The fitting to the algebraic decays for the solid and hexatic phases, and to the exponential decay for the liquid phase at long time scales are shown respectively, while for the short time scales algebraic decay is used for fitting the data. The KTHNY limit of the hexatic decay exponent $0.125$ is indicated by the solid green line.}}
\end{figure*}

The KTHNY theory predicts a constant $G_6(r)$ close to 1 in the solid phase. When entering the hexatic phase, $G_6(r)$ should decay algebraically $G_6(r) \propto r^{-\eta_6}$. The algebraic decay is slow and there is no characteristic length scale, thus the hexatic phase shows quasi-long range orientational correlations. A critical value of $\eta_6 \rightarrow 1/4$ is predicted by the KTHNY theory approaching the hexatic to liquid phase transition. In the liquid phase, exponential decay $G_6(r) \propto e^{-r/\xi_6}$, where $\xi_6$ is the orientational correlation length, is expected.

In Fig. 4A, $G_6(r)$ at different magnetic fields is shown in a $\log$-$\log$ coordinate system. Fits of the upper envelope of each $G_6(r)$ curve by power-laws is shown as dashed lines. An almost constant $G_6(r)$ close to 1 is obtained in the solid phase at $H=444$ Oe, demonstrating long range orientational order. In the hexatic phase, visible in the magnetic field ranging from 571 Oe to 1092 Oe, the algebraic decay manifests itself as linear decrease in the $\log$-$\log$ coordinate system. The decay exponents $\eta_6$ derived by fitting the data are shown in Fig. 4B. The largest value of $\eta_6$ obtained is 0.23 at $H=1092$ Oe, very close to the predicted critical value of 0.25 by the KTHNY theory (indicating by the solid green line in Fig. 4A). Further increasing the magnetic field to 1124 Oe results in a much fast decay that can be fitted by an exponential decay at large distances, indicating the complete loss of the orientational order. Note that an algebraic decay can still be evidenced in the liquid phase at short distances, with the decay exponent $\eta_6 = 0.76$, much larger than the critical value of 0.25. This originates from the existence of small hexatic islands in the liquid phase due to the spacial and temporal fluctuations.

Besides the spatial correlations, the temporal correlations of the local orientational order parameter $\Psi_6(\mathbf{r},t)$ is also revealed by analyzing LTEM movies acquired in the different phases. The temporal correlation function of the orientational order $G_6 (t)$ can be defined as\cite{scnmm_2018}:

\begin{eqnarray}
  \begin{aligned}
    &G_6 (t) = \left< \Psi_6 (\mathbf{r},\tau) \Psi_6^* (\mathbf{r},t+\tau)\right>_{\tau, \mathbf{r}} \\
  \end{aligned}
  \label{eq:eq8}
\end{eqnarray}
where the averages are taken over time and over the skyrmion positions. The results are shown in Fig. 4C, in which distinct behaviors of $G_6(t)$ can be observed. At low magnetic field, $H=571$ Oe, the temporal correlations remain essentially constant close to 1, qualitatively the same as observed for $G_6(r)$ in the solid phase. Algebraic decay is exhibited when increasing the magnetic field to $H=982$ Oe, demonstrating the hexatic phase behavior. For an even higher field, $H=1045$ Oe, similar to $G_6(r)$, short time scale power-law and long time scale exponential decays can be observed respectively, indicating a transition to the liquid phase with strong fluctuations.

The controllable realization of disordered, non-periodic configurations of skyrmion assemblies on top of a well organized atomic lattice is of extreme importance for the understanding of the particle nature of skyrmions and their interactions. While a simultaneous fading out of all the skyrmions in a SkL is expected by various simulations upon the increase of the system free energy, i.e., the vanishing of non-correlated individual skyrmions, our work demonstrates very complex decoherence processes prior to the destroying of skyrmions at high enough energy. The obtained insights may be extended to a large variety of interacting systems especially in 2D.

For applications, high information density and low energy dissipation should be fulfilled simultaneously in developing next generation magnetic storage techniques. Skyrmions are widely considered as promising candidates of information carriers, and the understanding of their collective behaviors in dense states are crucial. As such, our work thus not only provides a deep insight to the particle nature of skyrmions in bulk materials, but may also inspire pivotal progress towards skyrmion based spintronic technologies.

\section*{Acknowledgments}
We thank T. Giamarchi and A. Rosch for very insightful discussions and are grateful to D. Laub and B. B\'{a}rtov\'{a} for sample fabrication. \textbf{Funding:} This work was supported by the Swiss National Science Foundation (SNSF) through project 166298, the Sinergia network 171003 for Nanoskyrmionics, and the National Center for Competence in Research 157956 on Molecular Ultrafast Science and Technology (NCCR MUST), as well as the ERC project CONQUEST. \textbf{Author contributions:} P.H., F.C. and H.M.R. conceived the research, P.H. and M.C. designed the study, A.M. synthesized the crystalline samples, P.H. and M.C. performed the LTEM experiments, P.H., H.M.R. and F.C.analyzed the data, all authors contributed to the interpretation of the data. P.H., H.M.R. and F.C. wrote the paper. \textbf{Competing interests:} The authors declare no competing interests. \textbf{Data and materials availability:} All data needed to evaluate the conclusions in the paper are present in the paper and/or the supplementary materials.

\end{document}